\documentstyle[11pt,rtrpp4]{article}

\tighten

\received{}
\accepted{}
\journalid{}{}
\articleid{}{}
 
\slugcomment{Submitted to The Astronomical Journal}
 
\begin{document}
 
\title{Transformations between the theoretical and observational 
planes in the HST--NICMOS and WFPC2 photometric systems}
 
\author{Livia Origlia}
\affil{Osservatorio Astronomico di Bologna, 
        Via Ranzani 1, I--40127 Bologna, Italy\\
   e--mail: origlia@astbo3.bo.astro.it}
 
\and 

\author{Claus Leitherer}
\affil{Space Telescope Science Institute, 3700 San Martin Drive, Baltimore,
       MD 21218\\
       e--mail: leitherer@stsci.edu}

\begin{abstract}
 
Color--temperature relations and bolometric corrections in the 
HST--NICMOS F1110W, F160W and F222M and in the WFPC2 F439W, F555W and F814W 
photometric systems, using two different sets of model atmospheres, 
have been derived. 
This database of homogeneous, self--consistent 
transformations between the theoretical and observational planes also
allows combinations of visual and infrared quantities, without 
any further transformation between the two different photometric systems.
The behavior of the inferred quantities with varying the stellar parameters, 
the adopted model atmospheres and the instrumental configurations are
investigated.
Suitable relations to transform colors and bolometric corrections from 
HST to ground--based photometric systems are also provided.  

\end{abstract}
 
 
\section{Introduction}
 
Transformations between the theoretical and the observational planes 
are fundamental tools to compare stellar evolution models 
with observed color--magnitude diagrams.
In order to calibrate suitable relations among colors, temperatures and 
bolometric corrections (BCs) several basic ingredients are needed:
\begin{itemize}
\item homogeneous and complete grids of stellar 
spectra (observed and/or theoretical);
\item accurate filter profiles; 
\item reference spectra (the Sun, Vega etc.)
to set the zero points of the relations; 
\item suitable routines to interpolate within the grids  
and to integrate along the spectra.
\end{itemize}
These calibrations are certainly model dependent and systematic shifts 
between different scales are common features.
The goal of this paper is to obtain color--temperature relations and 
BCs in the HST--NICMOS photometric system. 
We also derive analogous transformations for a few selected filters in the 
WFPC2 system to provide homogeneous, self--consistent 
color--temperature relations and BCs  
when combinations of optical--infrared colors are used.
A more complete calibration of the color--temperature transformations 
in the WFPC2 system can be found in the paper by Holtzman et al. (1995, 
hereafter H95).

In Sect.~2 we describe the code we used to derive the transformations.
In Sect.~3 and 4 we discuss the transformations in the 
HST--NICMOS and the WFPC2 systems, respectively. 
In Sect.~5 we derive suitable relations to transform 
a few selected colors and BCs from the 
HST to the ground--based photometric system.
In Sect.~6 we draw our conclusions.

\section{The code}

The synthetic colors and BCs were computed 
using a modified version of the evolutionary synthesis code by 
Leitherer et al. (1999).

We implemented two different set of model atmospheres:  
{\it i)}~the compilation by Lejeune, Cuisinier \& Buser (1997, 
hereafter LCB97), 
based on the ATLAS code by Kurucz and corrected to match 
empirical color--temperature calibrations;  
{\it ii)}~the compilation  by Bessell, Castelli \& Plez 
(1998, hereafter BCP98) 
using only the homogeneous set of models without overshooting 
computed by Castelli (1997) and based on Kurucz's ATLAS9 models.
The grid of stellar parameters explored by our computations is:
\begin{itemize}
\item Metallicities in the range 0.01 -- 1.00 solar.
\item Effective temperatures $T_{\rm eff}$ in the range 3500 -- 50000 K.
\item Gravities log$~g$ in the range 0.0 -- 5.0. 
\end{itemize}

The ground--based BVIJHK filter profiles are in the Johnson's (1966) 
photometric system and were taken
from Buser \& Kurucz (1978) (BV filters) and from Bruzual (1983)
(IJHK filters) (see also Sect.~5 in Leitherer \& Heckman 1995 
for more details).
The selected broadband filters in the HST--WFPC2 and NICMOS 
photometric systems 
are:   F439W, F555W, F814W and F110W, F160W, F222M, respectively.
The filter profiles have been multiplied by the 
wavelength dependent detector quantum efficiency 
of the four WFPC2 and the three NICMOS cameras. 
We used the most updated post--launch throughput curves
and CCD quantum efficiencies, according to the
WFPC2 and NICMOS documentation on the WEB.
The quoted NICMOS throughput curves 
are already empirically adjusted to match standard star measurements
(the correction factors are less than 10\% 
in the case of the F110W and F160W filters and about 25\% for the F222M one).
We do not apply any further empirical adjustment to these
response curves.

The colors have been normalized assuming 0.00 values in all passbands 
for a Vega--like star with $T_{\rm eff}$=9500K, log$~g$=4.0 and [Fe/H]=--0.5 
(cf. e.g. BCP98).
The BCs have been normalized assuming a value of --0.07 
for a Sun--like star with $T_{\rm eff}$=5750~K, log$~g$=4.5 and [Fe/H]=0.0 
(cf. e.g. Montegriffo et al. 1998).
Different assumptions for the colors and bolometric corrections of
Vega and Sun --like stars can be accounted for by simply scaling the
inferred quantities by the corresponding amounts.

In order to quantify the influence of the adopted filter response curves
on the inferred colors we also compare our B--V and V--K values
with those in Table 2 of BCP98 for solar metallicity.
The adopted model atmospheres are exactly
the same and also similar within 0.01 mag are the reference colors for
the Vega--like star. The only difference between the two sets of 
transformations are the adopted filter response curves. 
Within 0.01 mag our V--K color is fully consistent with BCP98 values, while
only below 5000K our B--V becomes progressively bluer
(at most 0.03 mag at 3500K).

For each model atmosphere at a given $T_{\rm eff}$ and log$~g$ we 
tabulate the corresponding colors in the form (V-F), where F is the 
selected filter in the ground--based or in the HST system, 
and the bolometric correction BC$_V$ in the V passband.
We also provided analogous colors and BCs using simple 
blackbodies at a given temperature. 

All the tables with the computed transformations can be 
retrieved from http://www.stsci.edu/science/starburst/. 

In the following we analyze the behavior of the inferred quantities 
varying the stellar parameters and the adopted model atmospheres.  

\section{The HST--NICMOS infrared plane}

In Fig.~1 we plot the (F110W--F160W) and (F110W--F222M) color--temperature 
transformations in the HST--NIC2 system, using the BCP98 models 
for three gravities (log$~g$ of 0.0, 2.0 and 4.0) and two metallicities 
($Z=Z_{\odot}$ and $Z=0.1Z_{\odot}$).
For temperatures hotter than 4000~K these infrared colors show a 
scatter within 0.1~mag with varying log$~g$ and $Z$.
At lower temperatures the scatter among 
models with different gravities increases up to 
a few tenths of mag at $T_{\rm eff}$=3500~K, while the metallicity 
dependence is less critical. 

Pure blackbodies have systematically bluer colors than model atmospheres, 
especially at the coolest temperatures, as 
expected from the omission of molecular opacities. 
This means that for a given 
color, blackbodies are cooler than the model atmospheres.

Analogous color--temperature relations using the LCB97 models have been 
obtained.
In Fig.~2 we report the difference of the color--temperature 
transformations in the NIC2 system, by adopting the BCP98 or the 
LCB97 model atmospheres.
At temperatures below $\sim$5000~K the infrared colors using the LCB97 models 
get progressively bluer (up to 0.2--0.3~mag) 
than those using the BCP98 models.
This behavior  mainly reflects the difference between the 
original and corrected grids of model atmospheres by LCB97. 
Comparing their Figs.~6,7 (the original synthetic colors for giants and 
dwarfs, respectively) with their Figs.~14,15 
(the corresponding corrected models to match empirical color--temperature 
relations) one can see that in the range of temperatures between 4000 and 
3500~K the {\it corrected} (J--H) ground based color is bluer 
(from 0.1 up to 0.3 mag) than the corresponding original value, 
regardless the stellar gravity.
A similar comparison can be done between the original and 
the {\it corrected} (J--K) ground based color:  the latter becomes bluer 
with decreasing temperature by about 0.2 and up to 0.3 mag 
for giants and dwarfs at T$_{\rm eff}$=3500~K, respectively. 

The un--corrected LCB97 models, from the original grids of atmosphere 
spectra by Kurucz, should provide colors more similar to those 
obtained from the BCP98 models. 

Color--temperature transformations in the NIC1 
and NIC3 systems are also provided, as shown in Fig.~3. 
Regardess the model atmosphere used and the adopted stellar parameters, 
for a given temperature the (F110W--F160W) color in the NIC1 system is 
only $\le$0.01~mag redder than in the NIC2 system, while both the 
(F110W--F160W) and the (F110W--F222M) colors in the NIC3 system 
are slightly redder ($\le$0.03~mag below $\sim$4000~K). 

In Fig.~4 we plot the BC in the NIC2 
F110W, F160W and F222M 
passbands as a function of the temperature using the BCP98 models.
Increasing the temperature of the stellar atmospheres, 
progressively larger corrections 
(that is smaller values of the BC)
to the infrared fluxes have to be applied 
in order to get the bolometric luminosity.
The scatter in the inferred quantities for different gravities and 
metallicities is generally small.
At low temperatures the use of pure blackbodies requires 
larger corrections than using model atmospheres,
as expected since the former have bluer spectra than the latter.

In Fig.~5 we plot the difference between the values of the BC  
in the NIC2 photometric system, 
adopting the BCP98 or the LCB97 models, as we did for the colors in Fig.~2. 
At temperatures below 5000~K, larger BC values in the 
F110W and F222M and smaller ones in the F160W 
passbands (up to $\le$0.2~mag at $T_{\rm eff}$=3500~K) 
are required when LCB97 models are used compared to the BCP98 ones.

The above trends are almost independent of the adopted metallicity. 

For a given set of model atmospheres, 
very similar BCs within $<$0.01~mag are also obtained in the NIC1 and NIC3 
systems, as shown in Fig.~6. Only in the F110W passband we find that the BC          
values in the NIC3 system are slightly smaller ($\le$0.02~mag) than 
in the NIC2 system.

Our discussion on the behavior 
of the inferred colors and bolometric corrections 
with varying the stellar parameters
has been limited to the low temperature domain (T$_{\rm eff}<$5000~K) 
where they are more sensitive. 
At higher temperatures these infrared quantities 
become progressively less dependent on the adopted model atmospheres.

\section{The HST--WFPC2 visual plane}

In Fig.~7 we plot the (F439W--F555W) and (F555W--F814W) color--temperature 
transformations in the HST--PC1 system, using the BCP98 models 
for the three gravities and two metallicities of Fig.~1.

The scatter among models with different log$~g$ and $Z$ is
$\le$0.1~mag at all temperatures in the case of the 
(F555W--F814W) color, while 
at low temperatures ($T_{\rm eff}\le$5000~K)
the scatter among models with different gravities in the (F439W--F555W) color 
increases up to about 0.6~mag 
at $T_{\rm eff}$=3500~K, while the metallicity dependence is less critical.
As we found for the infrared colors, 
pure blackbodies show bluer colors than model atmospheres. 

Analogous color--temperature relations using the LCB97 models are also 
obtained.
In Fig.~8 we report the difference in the color--temperature 
transformations by adopting the BCP98 or the LCB97 model atmospheres.
Using the LCB97 models, at low temperatures the (F439W--F555W) color 
becomes progressively bluer (up to 0.1--0.2~mag) at low gravities 
and only slightly redder at larger ones, compared to the values 
obtained from the BCP98 models.   
The (F555W--F814W) color becomes rapidly redder (particularly at low gravities) 
than the corresponding quantities using the BCP98 models 
for T$_{\rm eff}\le$4000~K. This behavior has little 
metallicity dependence.
As for the infrared colors, this behavior  can be mainly ascribed to the 
corrections applied by LCB97 to the original model atmospheres.
Comparing again their Figs.~6,7 (the original synthetic colors) 
with their Figs.~14,15 (the corresponding corrected models) 
one can see that below 4000~K 
the {\it corrected} ground--based (B--V) color of giants becomes bluer  
and slightly redder for dwarfs, while the ground--based (V--I) color 
becomes rapidly redder (a few tenths of mag, even larger values for
giants). 

In Fig.~9 we plot the BC in the F439W, F555W and F814W 
passbands as a function of the temperature using the BCP98 models.
The required BC values using pure 
blackbodies are on average slightly larger than using model atmospheres, 
as expected since the former have bluer spectra than the latter 
and, contrary to the infrared case, 
in the visual range they tend to be brighter than the model atmospheres. 

In Fig.~10 we plot the difference between the values of the BC,
adopting the BCP98 or the LCB97 models, as we did in Fig.~8 for colors.
At temperatures below 5000~K, 
smaller BC values in the F439W and F555W and larger 
(up to $\le$0.2~mag at $T_{\rm eff}$=3500~K) 
in the F814W passbands are inferred  
if the LCB97 models are used. 
As for the infrared plane, the above trends are sensitive to 
the adopted gravity and almost 
independent of metallicity. 

For a given set of model atmospheres, very similar color--temperature
transformations and BC (within 0.01 mag) are obtained in all four WFPC2
cameras.

\section{Discussion}    

Model atmospheres are a powerful tool to calibrate suitable color--temperature 
transformations since homogeneous and 
complete grids for a wide range of stellar parameters are available.
Nevertheless, for some specific applications empirical scales 
even if they are less complete in terms of stellar parameters 
are preferred.

Using the BCP98 model atmospheres, we calibrated average
relations to transform several representative colors and 
BCs from the HST to the ground--based photometric system 
where most of the empirical scales are calibrated 
(cf. Montegriffo et al. 1998 and references therein). 
All the BCP98 models included in our grid of stellar parameters  
for all the three NICMOS and four WFPC2 cameras have been used to compute the 
best fits. 
Very similar best fit relations can be obtained using the LCB97 model
atmospheres.

In Fig.~11 the best model fits to the difference between the 
ground--based (J--H) and (J--K) colors 
and the NICMOS (F110W--F160W) and (F110W--F222M) ones, respectively,
as a function of the NICMOS quantities are shown.

The derived transformations are practically metallicity and gravity 
independent and 
very similar relations can be obtained adopting a particular metallicity or
gravity.

A cubic polynomial relation is required to transform the (F110W--F160W) 
into the ground--based (J--H) color, while a simple linear relation 
allows one to transform (F110W--F222M) into the ground--based (J--K) color.   
Very small ($<$0.01~mag) global {\it r.m.s} values have been obtained. 
The maximum scatter between the best fit 
and the model atmospheres occurs at the lowest temperatures 
($T_{\rm eff}$=3500--4000~K) and is $\le$0.06~mag in both colors.
 
For comparison, we also plot the colors of the standard stars 
with measured F110W, F160W and F222M and ground--based JHK magnitudes, 
according to the NICMOS Photometry Update WEB page (November 25, 1998), 
even though the quoted values are still in the process of being updated, 
and of a set of red stars in the Baade's window measured by Stephens et al. 
(1999).

Unfortunately, most of these stars are cooler than 3500K, that is out of the
temperature range covered by the selected set of model atmospheres. 
Nevertheless, the observed quantities are reasonably reproduced 
(within $\sim$0.1~mag) by the best model fits in the temperature 
range covered by the models, 
that is $T_{\rm eff}\ge$3500~K, while at lower temperatures 
the scatter is larger and also depends on the adopted extrapolation. 

In Fig.~12 we show the best model fits to the 
difference between the ground--based BC$_J$ and BC$_K$ 
and the corresponding NICMOS BC$_{F110W}$ and BC$_{F222M}$ values,
as a function of the (F110W--F160W) and (F110W--F222M) NICMOS colors, 
respectively.
Cubic polynomial relations with even smaller scatters than for colors 
are required to transform the BC from the NICMOS 
into the ground--based infrared photometric system.

The numerical relations to transform the selected colors and 
BCs from the NICMOS to the ground--based photometric system 
are listed below:

\noindent
$(J-H)=-0.063(F110W-F160W)^3+0.172(F110W-F160W)^2+0.563(F110W-F160W)+0.007$ 

\noindent
$(J-K)=0.803(F110W-F222M)+0.003$ 

\noindent
$BC_J=BC_{F110W}+0.069(F110W-F160W)^3-0.181(F110W-F160W)^2+
0.443(F110W-F160W)-0.008$ 

\noindent
$
BC_K=BC_{F222M}+0.023(F110W-F222M)^3-0.110(F110W-F222M)^2+
0.204(F110W-F222M)-0.001
$ 

In Fig.~13 the best model fits 
to the difference between the ground--based (B--V) and (V--I) colors 
and the corresponding WFPC2 (F439W--F555W) and (F555W--F814W) values 
as a function of the corresponding WFPC2 quantities are shown.   
Cubic polynomial relations are required to transform  
the (F439W--F555W) and the (F555W--F814W) colors into 
the corresponding ground--based (B--V) and (V--I) 
quantities. 

The global {\it r.m.s} values are still very small ($\le$0.02~mag) as 
for infrared colors, 
while the maximum scatter (which occurs at the lowest temperatures and 
reflects a gravity dependence in the case of the (B--V) and a metallicity 
dependence in the case of the (V--I)) 
between the best fit and model atmospheres with selected metallicity 
or gravity is somewhat larger 
($\sim$0.10--0.16~mag) than for the infrared colors.

A direct comparison between our color transformations in the (B--V) planes 
and those proposed by H95 using their Table 7 indicates an excellent 
agreement, with only a minor, systematic difference of
0.01 mag (our (B--V)--(F39W--F555W) are bluer than the
corresponding H95 values).

In Fig.~14 we report the best model fits to the 
difference between the ground--based BC$_V$ and BC$_I$ 
and the corresponding WFPC2 BC$_{F555W}$ and BC$_{F814M}$ values,
as a function of the (F439W--F555W) and (F555W--F814W) WFPC2 colors, 
respectively.
As for the corresponding infrared quantities, cubic polynomial relations 
with very small scatters 
are required to transform the BC from the WFPC2  
into the ground--based visual photometric system.

The numerical relations to transform the selected colors and 
BCs from the WFPC2 to the ground--based photometric system 
are listed below:

\noindent
$(B-V)=0.024(F439W-F555W)^3-0.136(F439W-F555W)^2+1.060(F439W-F555W)-0.014$

\noindent
$(V-I)=0.049(F555W-F814W)^3-0.077(F555W-F814W)^2+1.120(F555W-F814W)-0.005$

\noindent
$BC_V=BC_{F555W}+0.016(F439W-F555W)^3-0.070(F439W-F555W)^2+
2.092(F439W-F555W)-0.001$

\noindent
$BC_I=BC_{F814W}+0.052(F555W-F814W)^3-0.110(F555W-F814W)^2+
2.189(F555W-F814W)-0.007$

\section{Conclusions}    
 
Using our synthesis code we derive homogeneous, self consistent 
color--temperature relations and 
BCs in the HST infrared and visual photometric systems.

We investigated the behavior of the derived quantities for different 
stellar parameters and adopted set of model atmospheres.
Major results are:
\begin{itemize}
\item For a given set of model atmospheres, 
scatters larger than 0.01~mag in the 
inferred colors are only observed at low temperatures ($\le$6000~K) when 
models with different gravities and, to a lower degree, 
with different metallicities 
are compared. 
\item For given stellar parameters, the BCP98 and LCB97 sets of model atmospheres 
provide significantly different colors (up to a few tenths of a magnitudes) 
below 5000~K. The inferred discrepancies can be mainly ascribed to the 
corrections applied by LCB97 to the original grids 
of Kurucz's model atmosphere spectra  
in order to match empirical color--temperature 
relations. 
\item The inferred BCs are less dependent on the 
adopted stellar parameters and model atmospheres than 
the colors, even at low temperatures.
\item Very similar colors and BCs have 
been obtained for the different NICMOS and WFPC2 cameras. 
\item Average relations, 
with a negligible dependence on the stellar parameters and the selected 
NICMOS or WFPC2 cameras, can be adopted to provide useful  
transformations between the HST and ground--based photometric systems. 
In the low temperature domain the ground--based (B--V), (J--H) and (J--K) 
colors are bluer than the corresponding ones in the HST photometric system, 
while the ground--based (V--I) color is redder than the corresponding 
(F555W--F814W) one in the WFPC2 system. 

\end{itemize} 

\acknowledgments

We acknowledge Jon Holtzman for his careful Referee Report and comments
and Laura Greggio for the helpful discussions and suggestions.
We thank the STScI NICMOS Staff for providing the information on the 
NICMOS throughput curves. 
L.O. acknowledges the financial support of the 
{\it ``Ministero della Universit\`a e della Ricerca
Scientifica e Tecnologica''} (MURST) to the project {\it Stellar Evolution}.

\clearpage

\begin{figure*}
\vskip6.9truein
\includegraphics{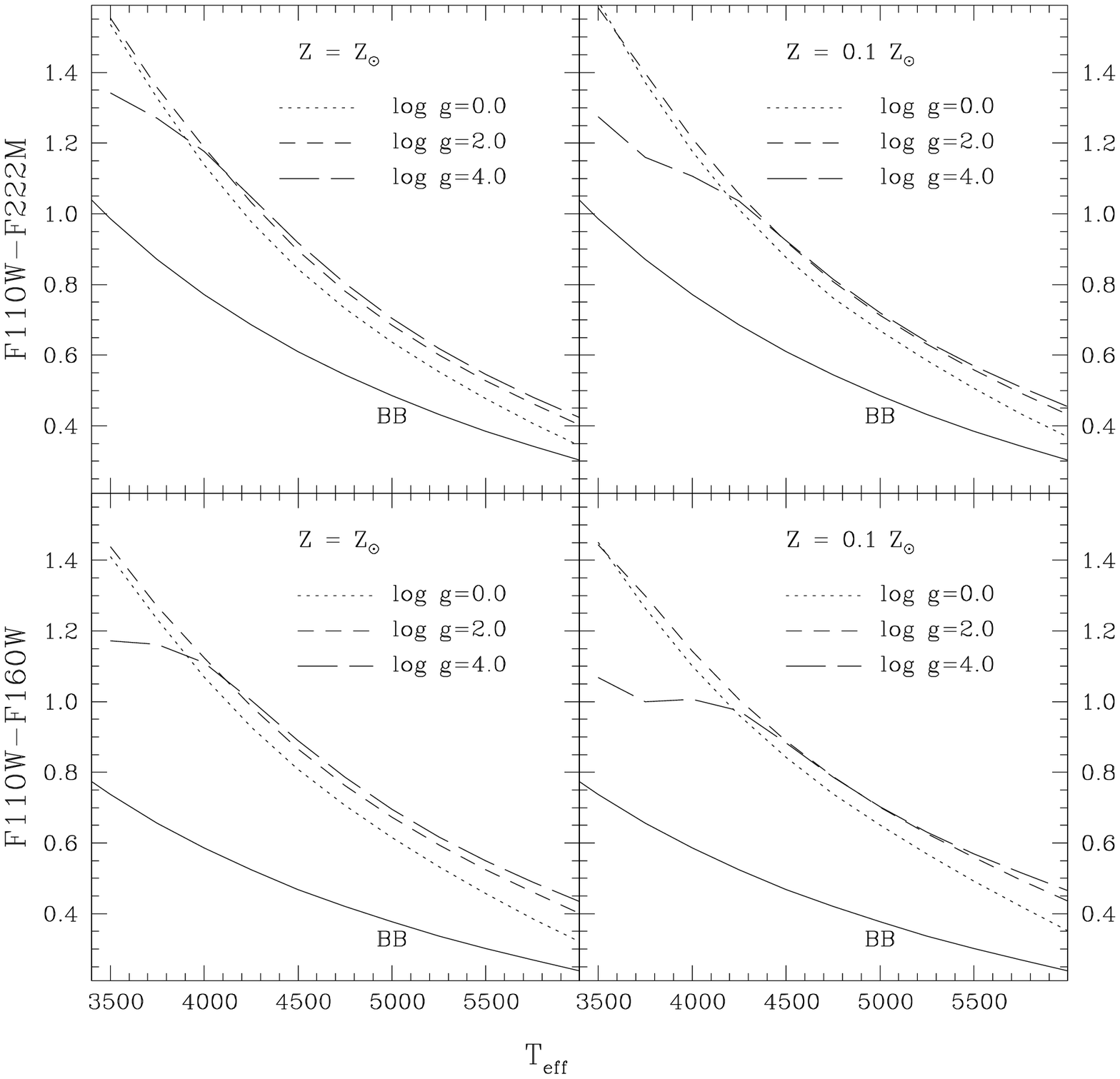}
\figcaption{
\label{fig1} 
(F110W--F160W) (bottom panels) and (F110W--F222M) (top panels) 
color--temperature
transformations in the HST--NIC2 system, using the BCP98 models.
Relations using three different gravities 
(log$~g$=0.0: dotted line, log$~g$=2.0: short--dashed line; log$~g$=4.0:
long--dashed line) and two metallicities
($Z=Z_{\odot}$ left panels, $Z=0.1Z_{\odot}$ right panels) are 
shown. For comparison we also plot the relations for a pure 
blackbody (continuous line). 
}
\end{figure*}
\clearpage

\begin{figure*}
\vskip6.9truein
\includegraphics{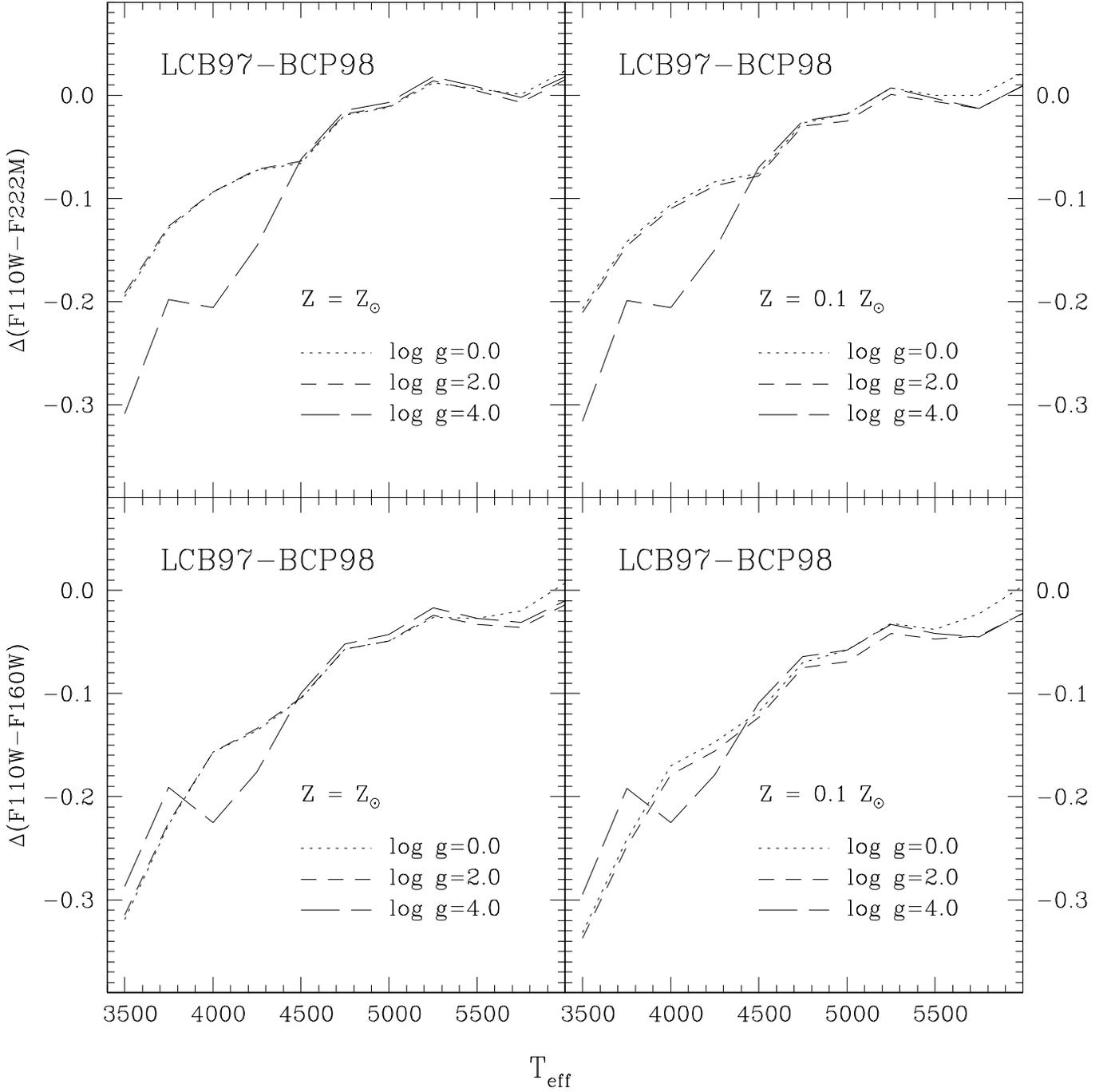}
\figcaption{
\label{fig2} 
Differences of the color--temperature
transformations in the NIC2 system, by adopting the BCP98 or the
LCB97 model atmospheres.
Notation as in Fig.~1.
}
\end{figure*}
\clearpage

\begin{figure*}
\vskip6.9truein
\includegraphics{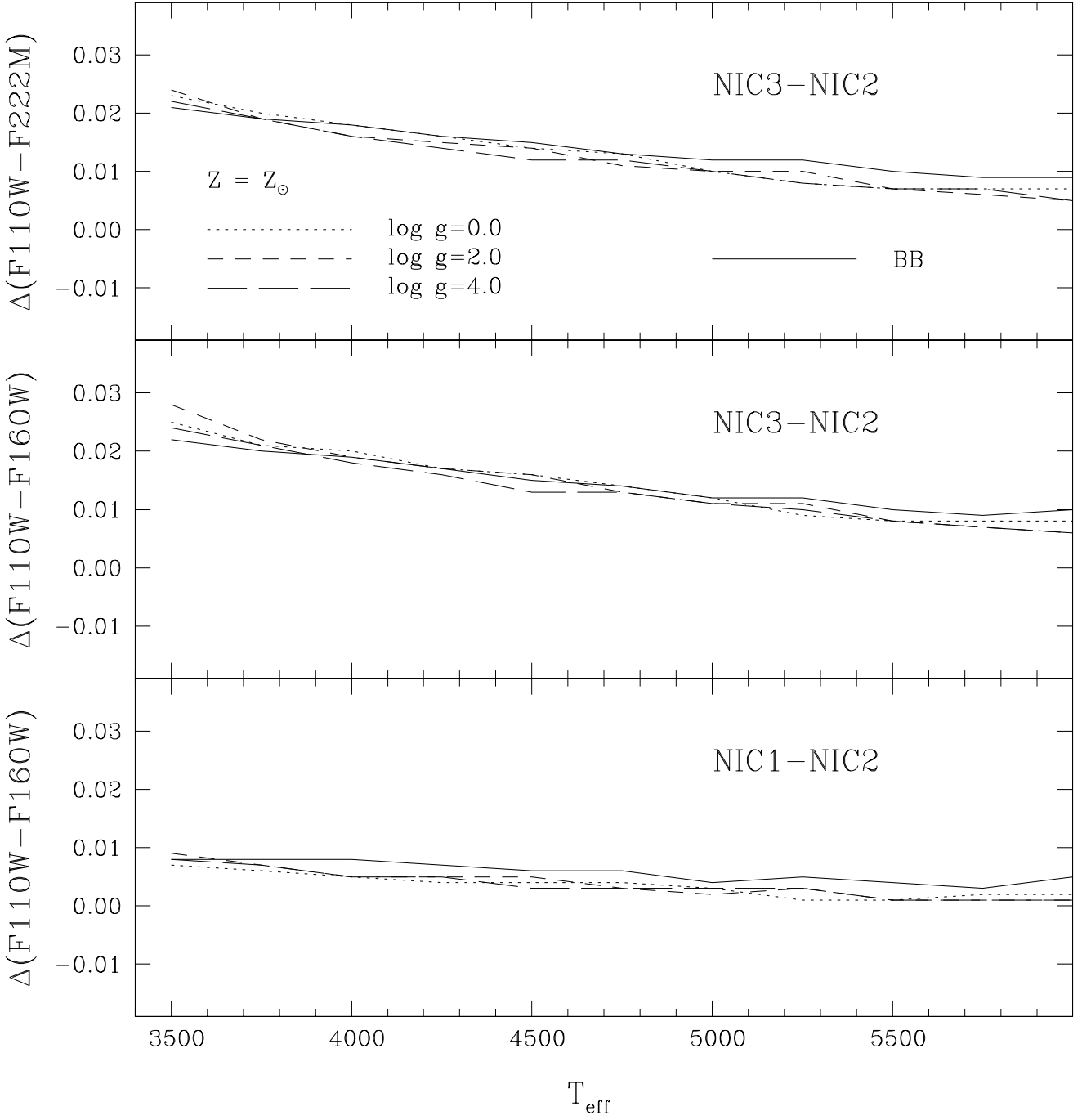}
\figcaption{
\label{fig3} 
Differences of the (F110W--F160W) and (F110W--F222M) color--temperature
transformations using the BCP98 models between the NIC2 and NIC1 or NIC3 
systems at solar metallicity. Notation as in Fig.~1.
}
\end{figure*}
\clearpage

\begin{figure*}
\vskip6.9truein
\includegraphics{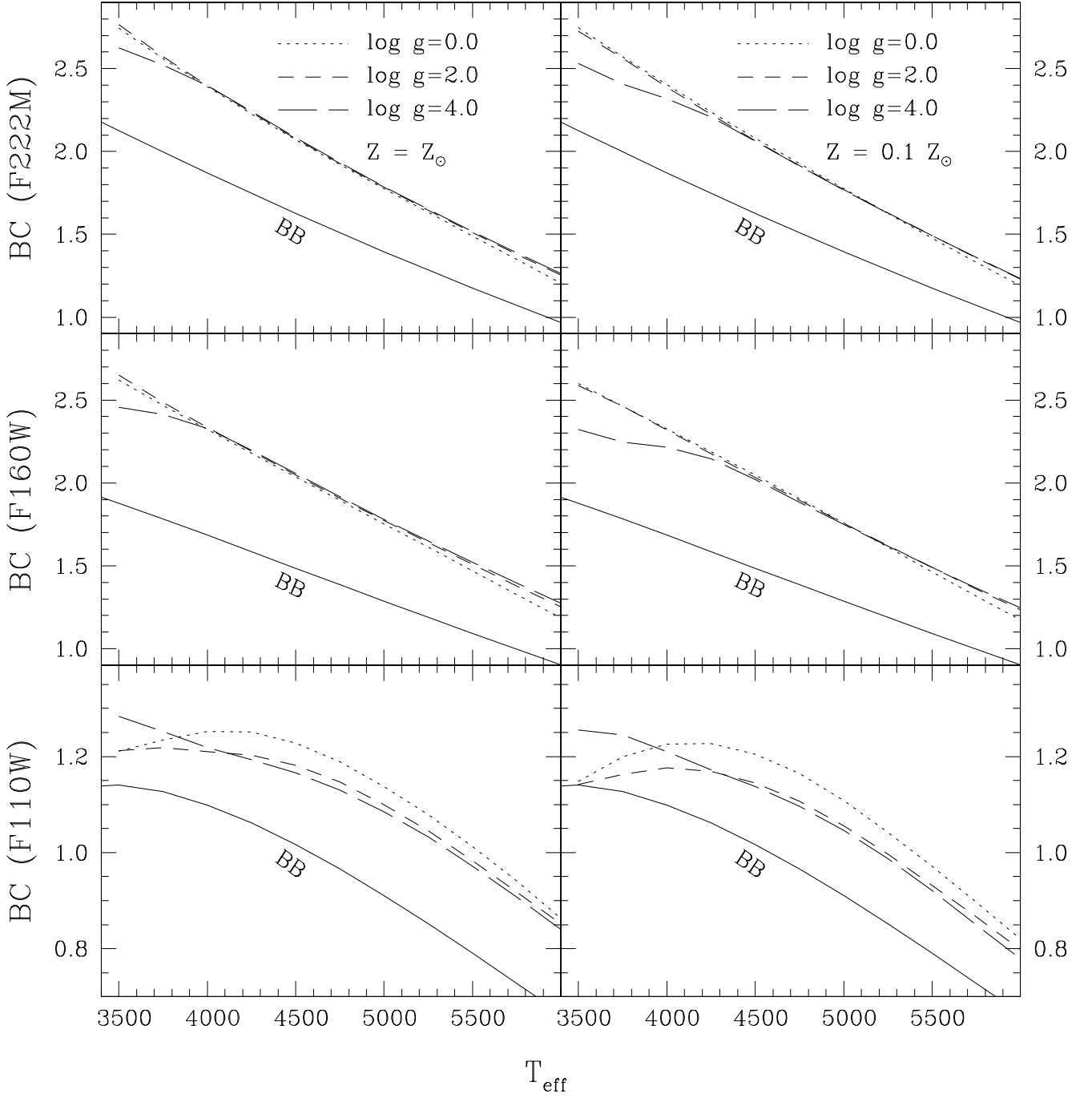}
\figcaption{
\label{fig4} 
Bolometric corrections in the NIC2
F110W, F160W and F222M
passbands as a function of temperature using the BCP98 models.
Notation as in Fig.~1.
}
\end{figure*}
\clearpage

\begin{figure*}
\vskip6.9truein
\includegraphics{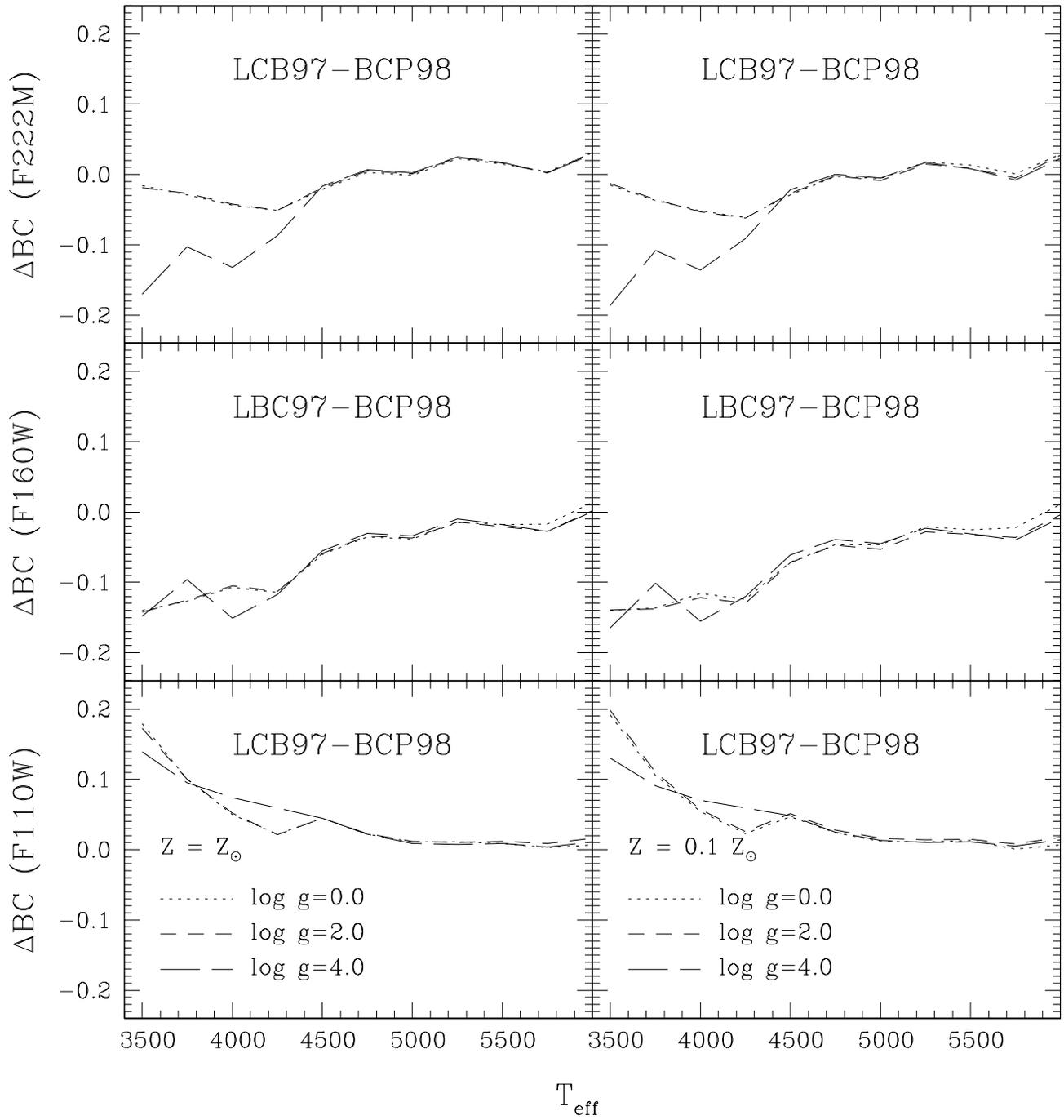}
\figcaption{
\label{fig5} 
Differences between the values of the bolometric
corrections in the NIC2 photometric system,
adopting the BCP98 or the LCB97 models. 
Notation as in Fig.~1.
}
\end{figure*}
\clearpage

\begin{figure*}
\vskip6.9truein
\includegraphics{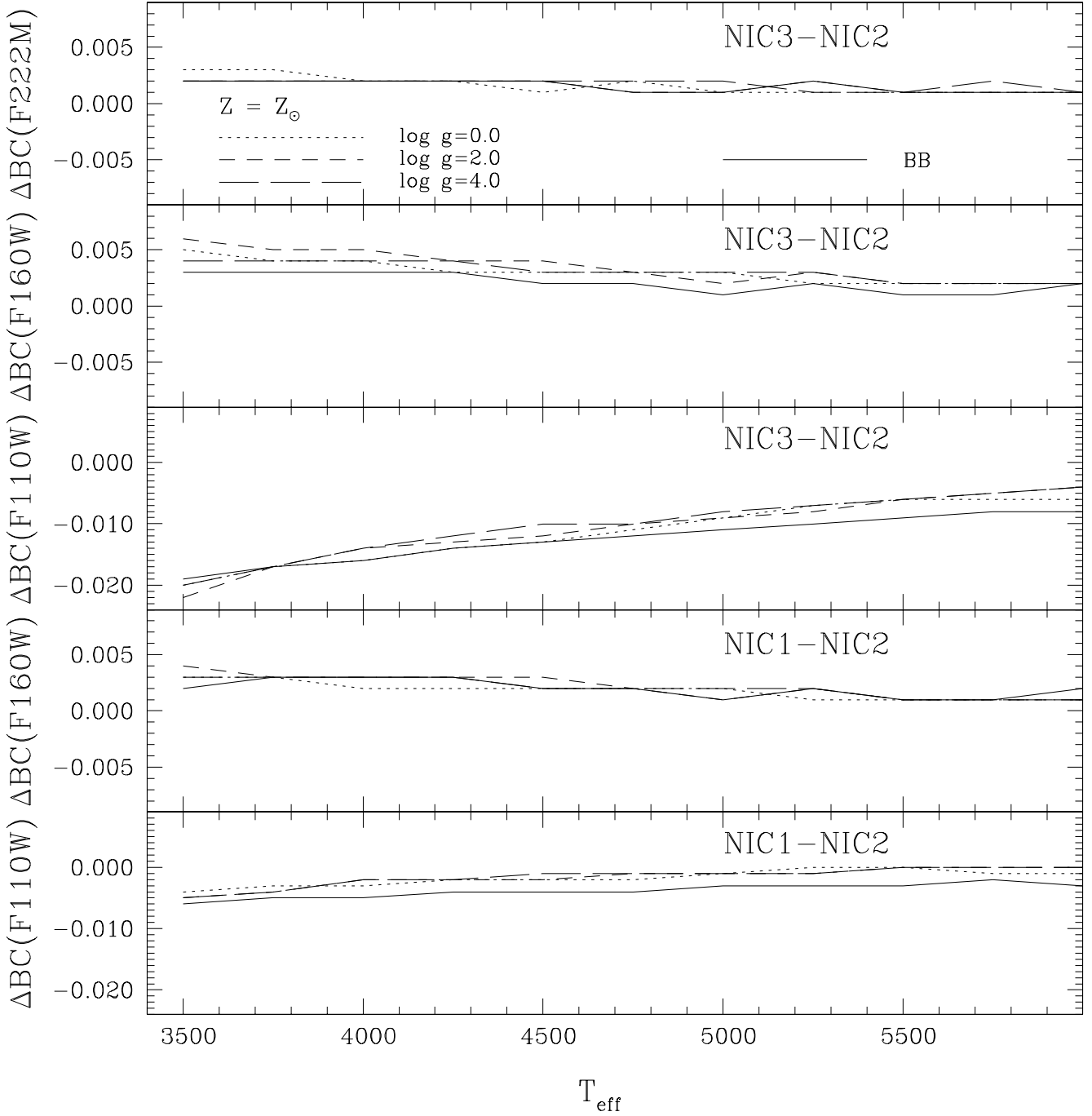}
\figcaption{
\label{fig6} 
Differences of the F110W, F160W and F222M bolometric corrections 
using the BCP98 models between the NIC2 and NIC1 or NIC3 systems at 
solar metallicity. Notation as in Fig.~1.
}
\end{figure*}
\clearpage

\begin{figure*}
\vskip6.9truein
\includegraphics{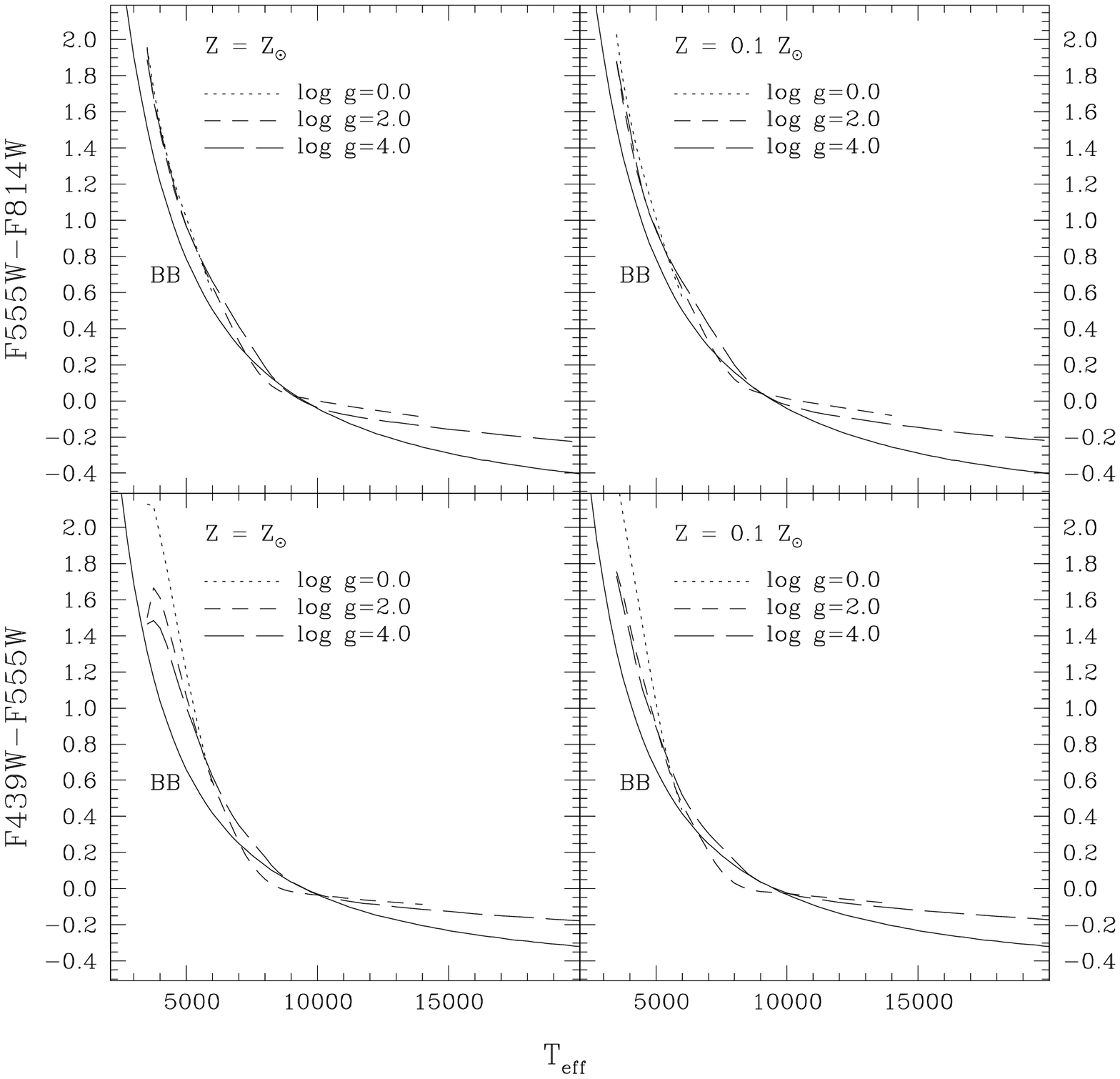}
\figcaption{
\label{fig7} 
As in Fig.~1 but for the (F439W--F555W) and (F555W--F814W) colors 
in the PC1 system.
}
\end{figure*}
\clearpage

\begin{figure*}
\vskip6.9truein
\includegraphics{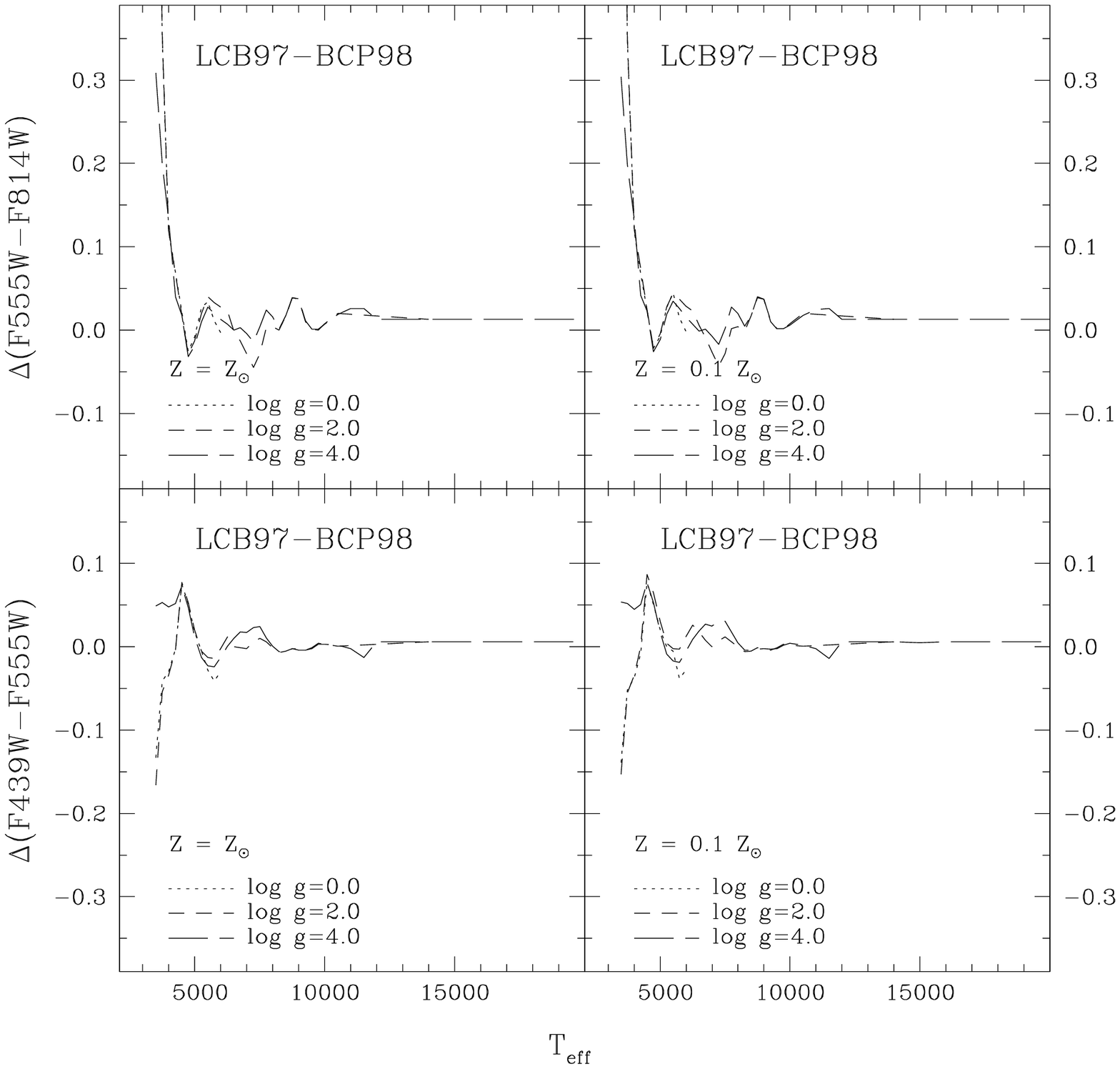}
\figcaption{
\label{fig8} 
As in Fig.~2 but for the (F439W--F555W) and (F555W--F814W) colors 
in the PC1 system.
}
\end{figure*}
\clearpage

\begin{figure*}
\vskip6.9truein
\includegraphics{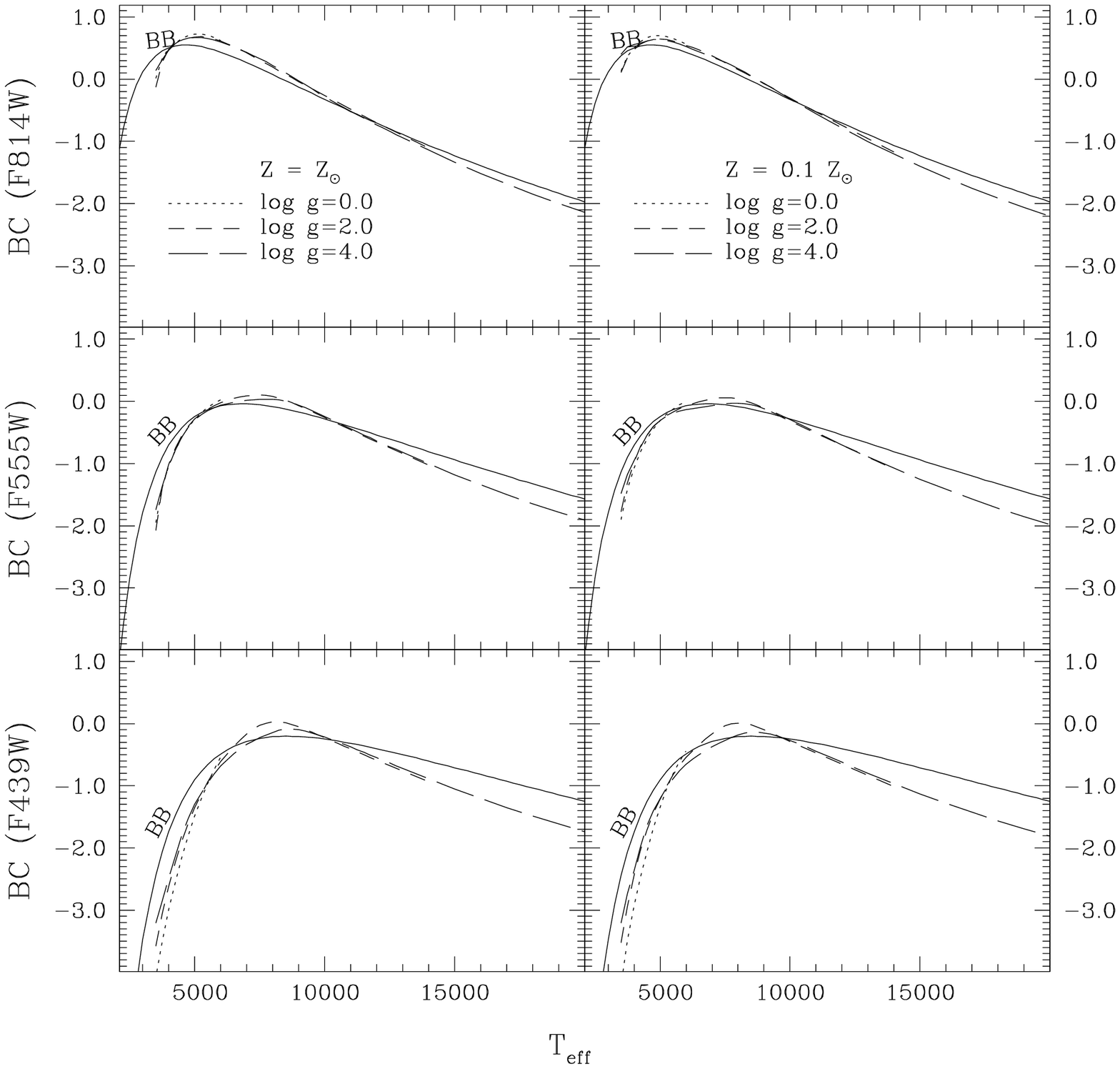}
\figcaption{
\label{fig9} 
As in Fig.~4 but for the F439W, F555W and F814W bolometric 
corrections in the PC1 system.
}
\end{figure*}
\clearpage

\begin{figure*}
\vskip6.9truein
\includegraphics{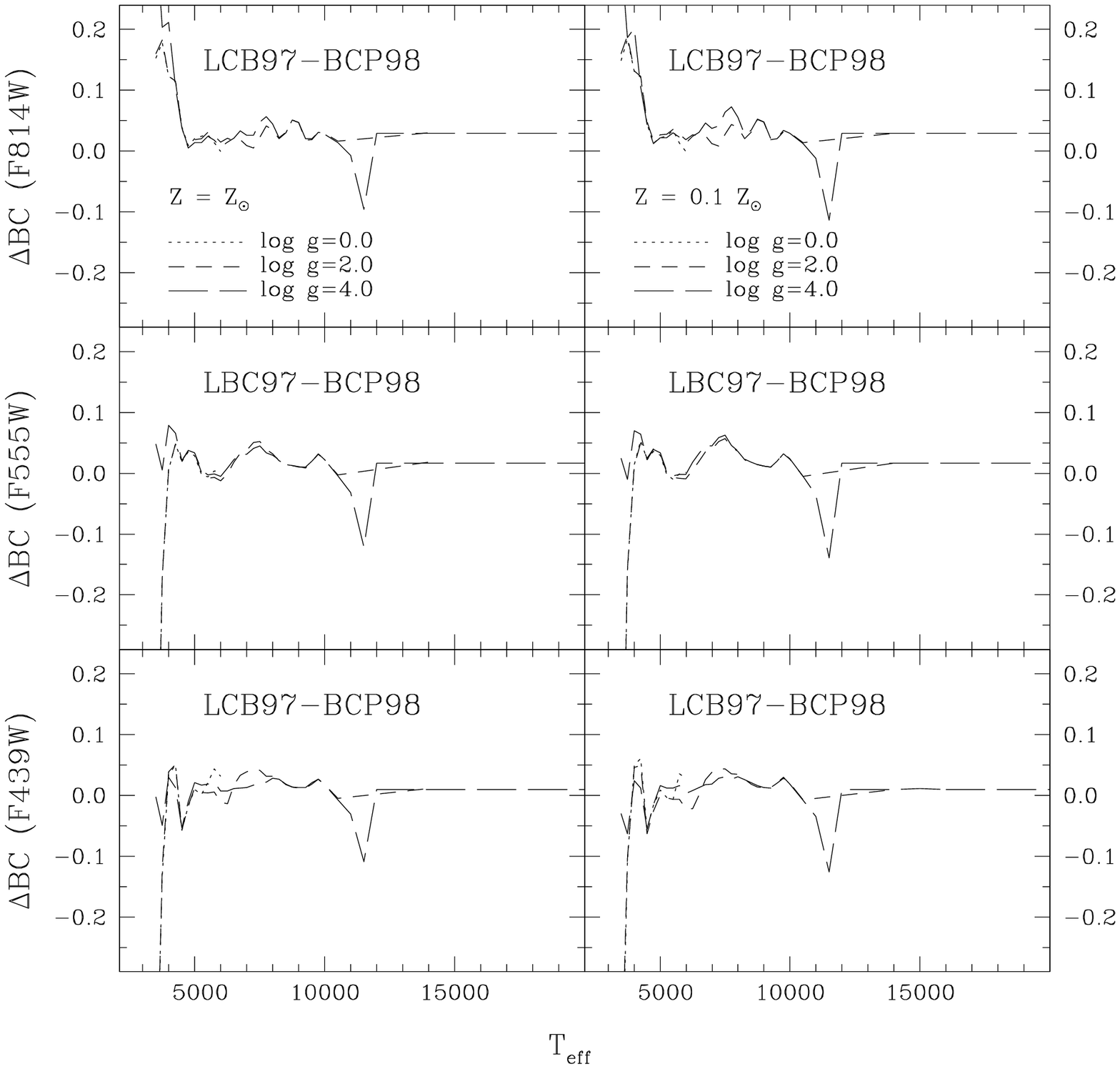}
\figcaption{
\label{fig10} 
As in Fig.~5 but for the F439W, F555W and F814W bolometric 
corrections in the PC1 system.
}
\end{figure*}
\clearpage

\begin{figure*}
\vskip6.9truein
\includegraphics{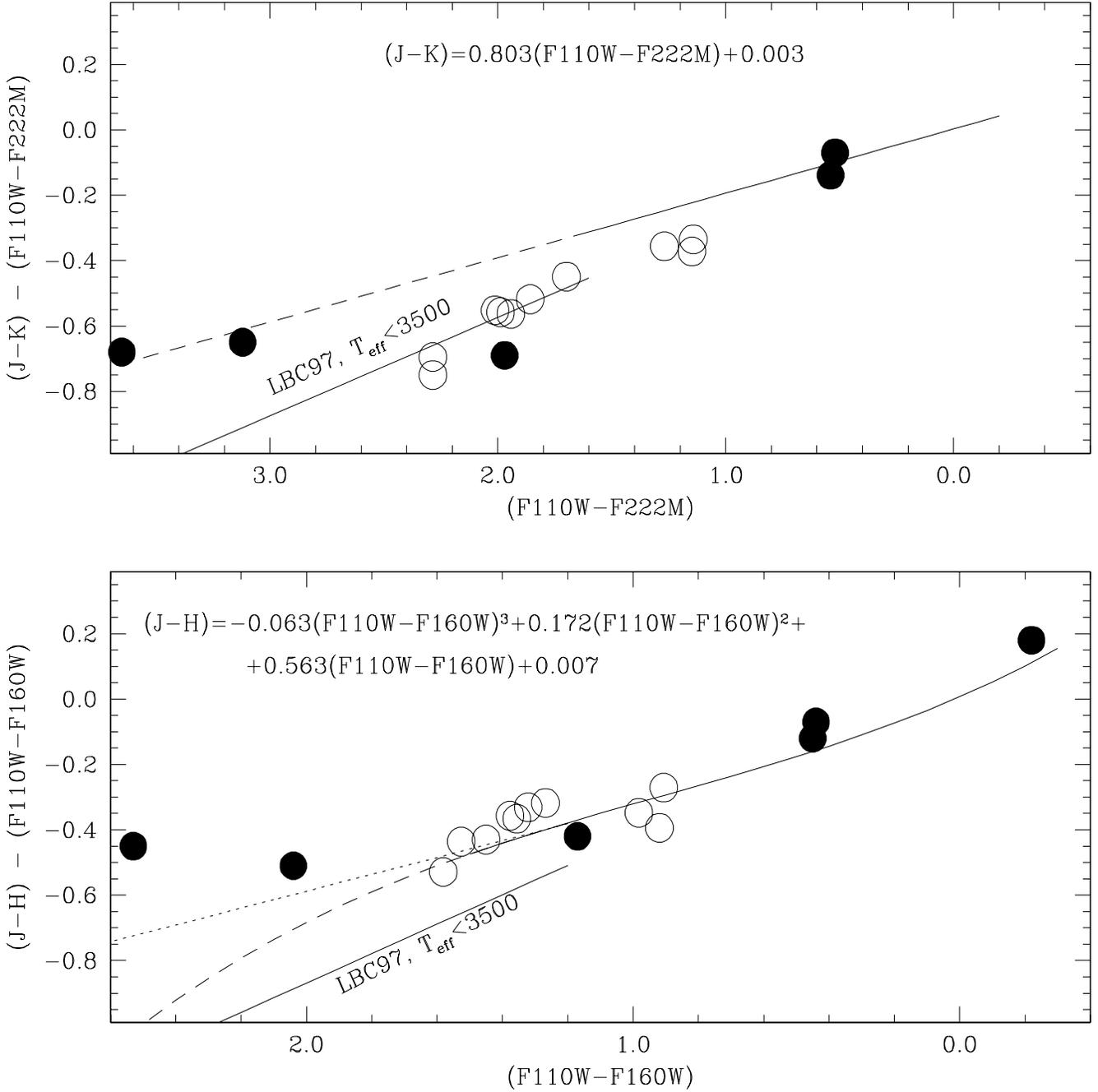}
\figcaption{
\label{fig11} 
Model best fits 
to the difference between the (J--H) and (J--K) colors in the ground--based 
photometric system and the 
corresponding (F110W--F160W) and (F110W--F222M) values in 
the NICMOS system, 
as a function of the corresponding NICMOS quantities.
Atmosphere: BCP98. 
Full circles are the observed values of the standard stars, according to 
the NICMOS Photometry Update WEB page.
Empty circles refer to the 
set of red stars in the Baade's window measured by Stephens et al. (1999) 
with photometric errors $\le$0.05 mag.
Dotted and shaded lines are the linear and cubic extrapolated relations, 
respectively. 
For comparison, we also plotted the best fit using the LCB97 model 
atmospheres at temperatures below 3500~K. 
}
\end{figure*}
\clearpage

\begin{figure*}
\vskip6.9truein
\includegraphics{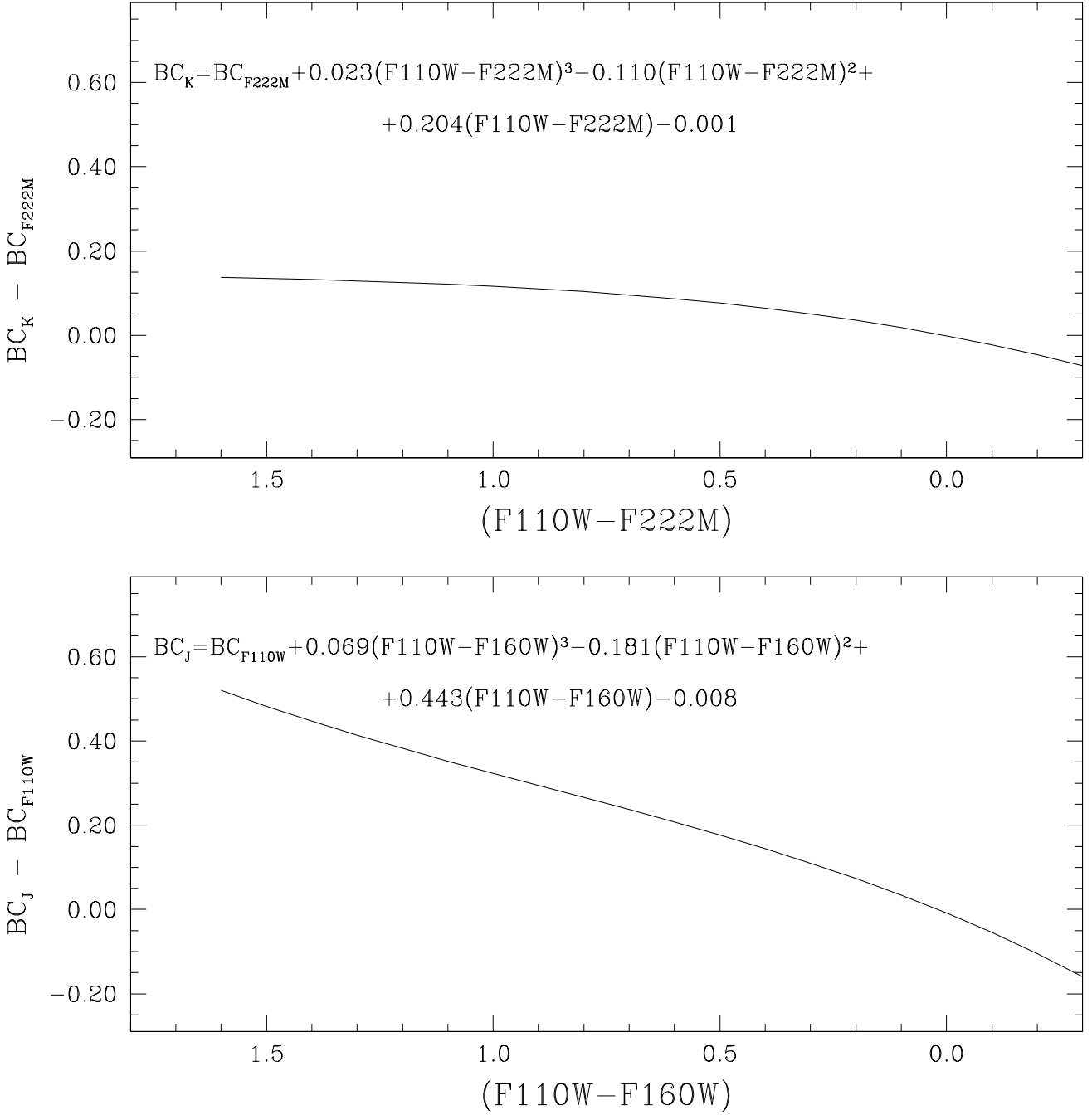}
\figcaption{
\label{fig12} 
Model best fits 
to the difference between the BC$_J$ and BC$_K$ bolometric corrections in 
the ground--based 
photometric system and the corresponding BC$_{F110W}$ and BC$_{F222M}$ values 
in the NICMOS system,
as a function of the (F110W--F160W) and (F110W--F222M) NICMOS colors, 
respectively.
Atmosphere: BCP98. 
}
\end{figure*}
\clearpage

\begin{figure*}
\vskip6.9truein
\includegraphics{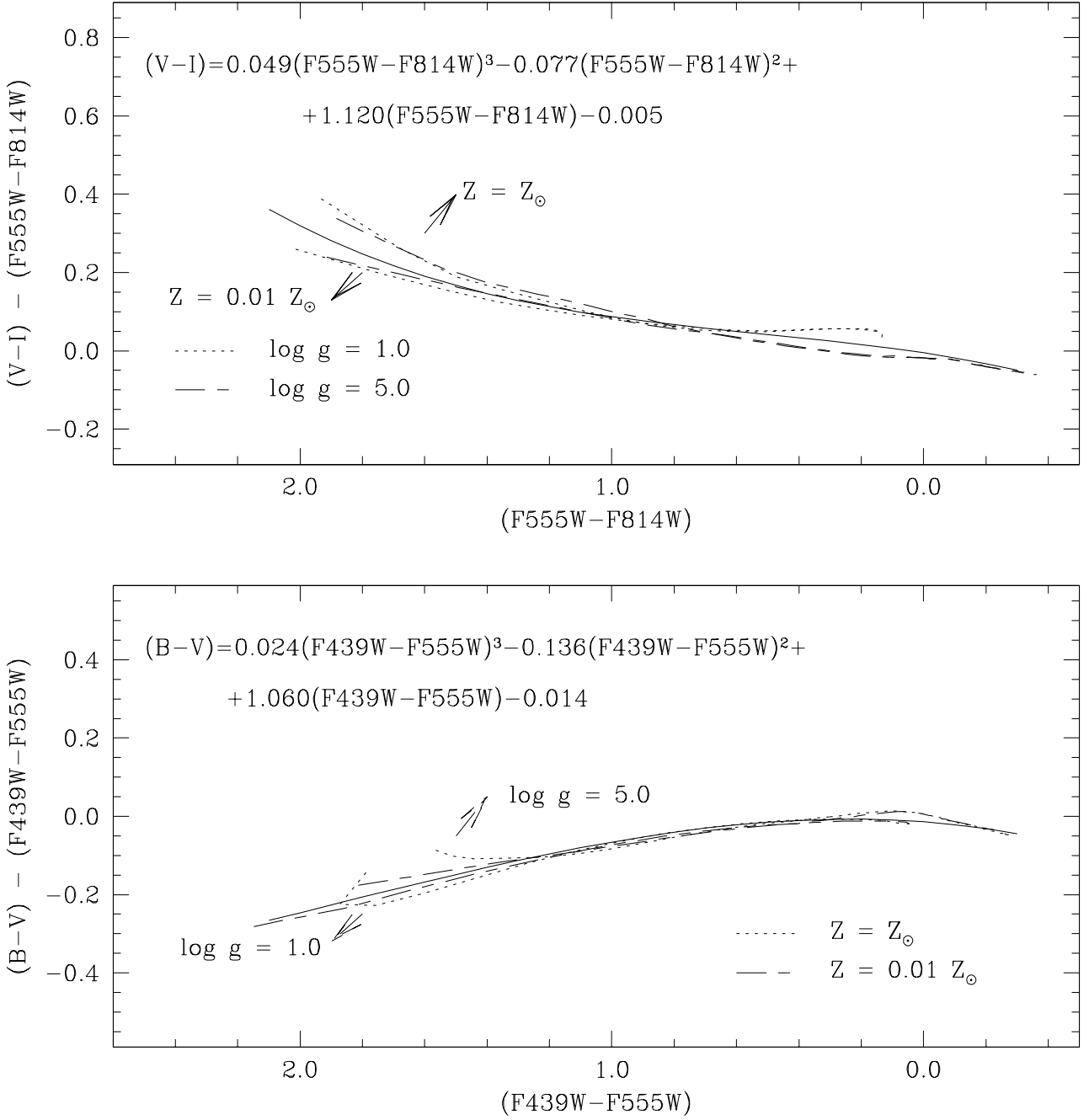}
\figcaption{
\label{fig13}
Model best fits 
to the difference between the (B--V) and (V--I) colors in the ground--based 
photometric system and the 
corresponding (F439W--F555W) and (F555W--F814W) values in 
the WFPC2 system, 
as a function of the corresponding NICMOS quantities.
Atmosphere: BCP98. 
For comparison, together with the best fit 
we also plot the models at the two different gravities 
(log$~g$=1.0 and 5.0) and metallicities ($Z=Z_{\odot}$ and $Z=0.01Z_{\odot}$).
}
\end{figure*}
\clearpage

\begin{figure*}
\vskip6.9truein
\includegraphics{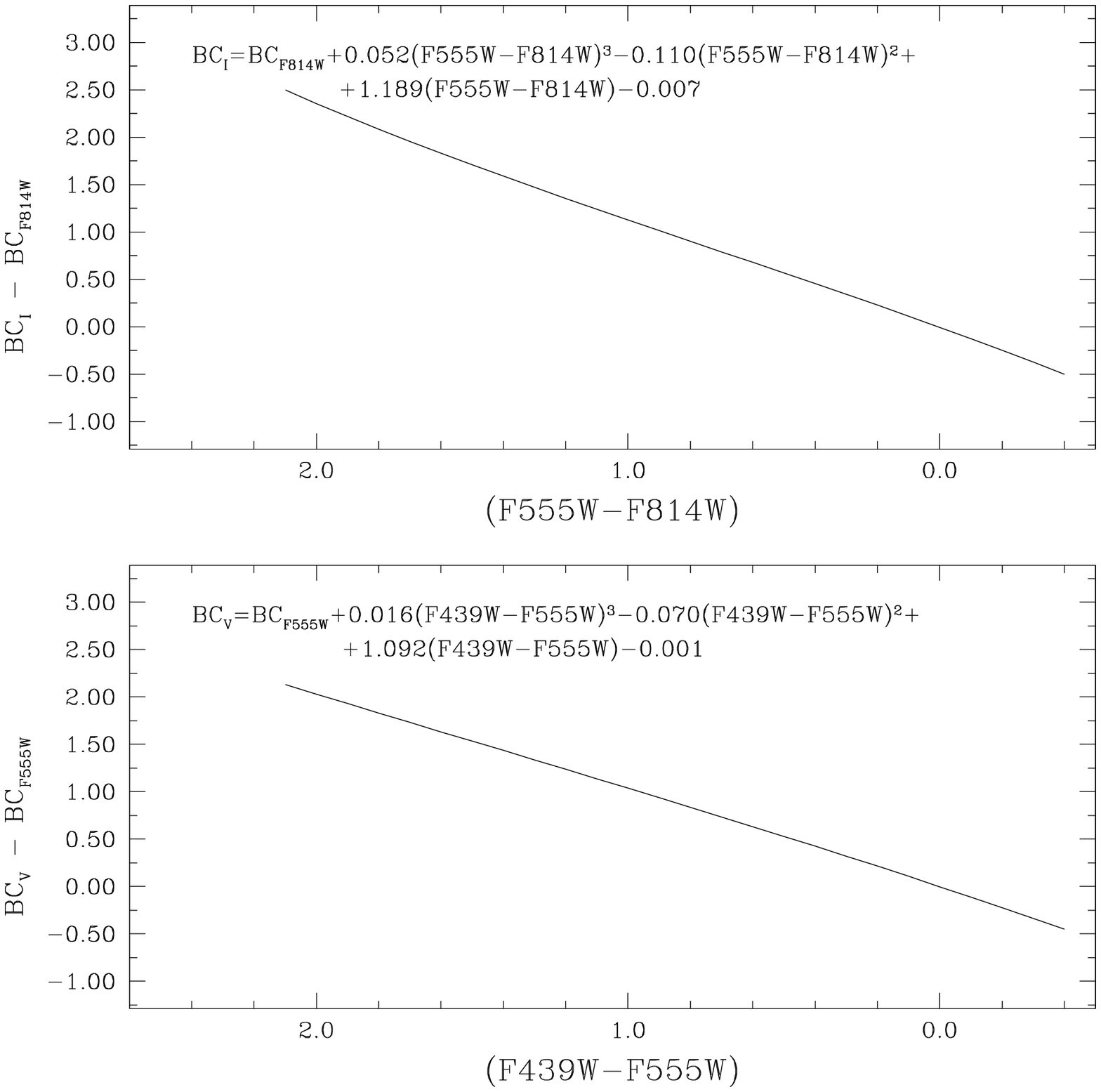}
\figcaption{
\label{fig14} 
Model best fits 
to the difference between the BC$_V$ and BC$_I$ bolometric corrections in the 
ground--based 
photometric system and the corresponding BC$_{F555W}$ and BC$_{F814M}$ values 
in the HST system,
as a function of the (F439W--F555W) and (F555W--F814W) WFPC2 colors, 
respectively.
Atmosphere: BCP98. 
}
\end{figure*}
\clearpage


\begin{references}

\reference{} Bessel, M. S., Castelli, F., \& Plez, B. 1998, \aap, 333, 231 
(BCP98)

\reference{} Castelli, F., 1997, private communication

\reference{} Holtzman, J. A., Burrows, J., Casertano, S., 
Hester J., Trauger, J. T., Watson, A. M., \& Worthey, G. 
1995, \pasp, 107, 1065 (H95)

\reference{} Johnson, H. L. 1966, \araa, 4, 193

\reference{} Leitherer, C., \& Heckman, T. M. 1995, \apjs, 96, 9 

\reference{} Leitherer, C., et al. 1999, \apjs, in press  

\reference{} Lejeune, T., Cuisinier, F., \& Buser, R. 1997, \aaps, 125, 229 
(LCB97)

\reference{} Montegriffo, P., Ferraro, F.R., Origlia, L., \& Fusi Pecci, F. 
1998, \mnras, 297, 872

\reference{} Stephens, W., Frogel, J.A., Ortolani, S., Davies, R., 
Jablonka, P., \& Renzini, A., 1999, astro-ph/9909001

\end{references}
\end{document}